\documentstyle[12pt]{article}

\newcommand{\be}{\begin{equation}}
\newcommand{\ee}{\end{equation}}
\newcommand{\ba}{\begin{eqnarray}}
\newcommand{\ea}{\end{eqnarray}}
\newcommand{\nl}{\newline}
\newcommand{\ft}{\footnote}
\newcommand{\w}{\Omega}
\newcommand{\al}{\alpha}
\newcommand{\bt}{\beta}
\newcommand{\ga}{\gamma}
\newcommand{\rt}{\rightarrow}
\newcommand{\Th}{\Theta}
\newcommand{\Ph}{\Phi}
\newcommand{\sg}{\sigma}

\begin{document}

\begin{flushright}
QMW-PH-96-27
\end{flushright}

\begin{center}

\Large{\bf Dirichlet Joyce Manifolds, Discrete Torsion\nl
 and Duality}

B.S.Acharya\ft{e-mail:r.acharya@qmw.ac.uk\nl
Work supported by PPARC}
{\it Queen Mary and Westfield College, Mile End Rd., London. E1 4NS}

\end{center}

\begin{abstract}

Using $U$-duality transformations we map perturbative Type IIA
string theory compactified on a class of Joyce $7$-manifolds
to a $D$-strings on $D$-manifold description in Type IIB theory.
For perturbative Type IIB theory on the same class of Joyce
manifolds we use duality transformations to map to an $M$-theory,
$M$-manifold
description, which is an orientifold with fivebrane twisted sectors.
$D$ and $M$-manifold analogues of Joyce orbifolds with discrete torsion
are found.
For the same class of compactifications we
show that Type IIA/IIB theory on a Joyce orbifold without (with)
discrete torsion is $T$-dual to Type IIB/IIA theory on
the same orbifold with (without) discrete torsion. For
this class of Type II compactifications this
proves an extension of the Papadopoulos-Townsend
conjecture, which states that the Type IIA and IIB theories
compactified on the same Joyce $7$-manifold are equivalent.
Finally we note that the Papadopoulos-Townsend conjecture
is a special case of the Generalised Mirror Conjecture.

\end{abstract}
\newpage

\section{Introduction}

$U$-duality, \cite{U}, interchanges perturbative string states 
(for example standard
twisted sector states of an orbifold compactification) with $D$-branes 
\cite{Pol} and $M$-branes \cite{Mbrane}. For example, in the case of Type IIA
on $K3$, $U$-duality transforms the ``standard'' perturbative description 
into a Type IIB, $D$-strings on $D$-manifold description \cite{Kut,D,Op}. For
the simple case when $K3$ is the $Z_2$ orbifold of $T^4$, $U$-duality 
exchanges the twisted sector IIA states with configurations
of $RR$ $D$-fivebranes in the IIB theory \cite{Kut}. For the case of Type IIB 
on ${T^4}/{Z_2}$, the perturbative compactification can be transformed to 
an orientifold of $M$-theory \cite{mu,w2}, in which $M$-fivebranes carry the
duals of the twisted sector states \cite{w2}\ft{For convenience, we
will refer to such backgrounds of $M$-theory as $M$-manifolds.}.

In this sense, $D$-brane and $M$-brane configurations encode information
about $U$-dual compactification spaces\ft{For instance, couplings in a
compactified theory are determined by topological intersection numbers of
the compactifying space. In principle, these should be determinable in
the dual $D$ or $M$-manifold description.}. The relationship between 
branes, geometry and physics has been investigated in a variety of
fascinating contexts in recent months (see for example \cite{...}).

The purpose of this note is twofold. The first is to exploit the analogy
between the construction of $K3$ as a blown up
toroidal orbifold and the construction
of manifolds of exceptional holonomy using the same technique \cite{J1,J2,J3}. 
This will give us a simple generalisation of the $D/M$-manifold
descriptions of Type IIA/IIB on ${T^4}/{Z_2}$, to the case of Joyce 
$7$-manifolds of $G_2$ holonomy. In doing this we
will also find further examples
of $D$ and $M$-manifold 
descriptions of discrete torsion, as have been found recently in 
\cite{Dtor,Go}.
We find that the geometry of the Joyce manifold compactification is, 
as expected, beautifully encapsulated by the fivebrane twisted sectors.
We present these constructions in sections two and three. These results
are evidence that Joyce 7-manifolds have points in their moduli
space in which $A_n$ singularities appear.

The second purpose of this note is to explore what may loosely be called 
``mirror conjectures'' for Type II compactifications on Joyce $7$-manifolds.
In \cite{shat}, the conformal field theory description of spaces of
exceptional holonomy was given. In order to interpret certain results of
that description, Shatashvilli and Vafa proposed a Generalised Mirror
Conjecture, which is proposed to apply to any quantum sigma model. Essentially
this conjecture states that if any ambiguity arises in the determination
of topological properties of the target space using quantum field theory, then
there exists a dual quantum field theory which resolves the ambiguity. A
classic example is given by the Witten index, $Tr{(-)^F}$. In 2d supersymmetric
sigma models,
this index computes the Euler character of the target space, up to
an overall sign ambiguity \cite{con}. 
Thus, in this case, the Shatashvilli-Vafa (SV)
conjecture requires the existence of two dual sigma models, which differ by a
sign in the calculation of the Witten index on supersymmetric ground states. 

On the other hand, compactification of both the Type IIA and IIB theories on
a general $7$-manifold of $G_2$ holonomy, $J$, 
was considered in \cite{pap}. There it was observed that
the massless spectra of the two 3d theories agreed (after dualising all 
vectors to scalars), and it was conjectured on that basis that both the IIA
and IIB theories compactified on the same $G_2$ $7$-manifold are physically
equivalent, or dual. We will refer to this conjecture as the PT conjecture.
The PT conjecture, as stated in \cite{pap}, applies to an arbitrary manifold
of $G_2$ holonomy. In this sense, it applies to IIA/IIB backgrounds which
are well defined in classical geometry. However, because such
backgrounds admit a description as conformal field theories, it is
natural to propose that the PT conjecture applies to ``$G_2$'' backgrounds
which do not have an interpretation in classical geometry, but are well
defined in CFT.

However, in this paper, we will restrict our attention to the largest
class of manifolds constructed in \cite{J2}. Specifically we
will consider manifolds of $G_2$ holonomy constructed as desingularisations
of ${T^7}/{{Z_2}^n}$. In principle, there may exist orbifolds in this class
which admit no desingularisation in classical geometry and our analysis
applies to these cases as well; but as examples we will only consider cases
considered by Joyce \cite{J2}.

For the ${T^7}/{{Z_2}^n}$ orbifolds, it turns out (following \cite{dt}) 
that we will be able
to prove an extension of the PT conjecture. We will show that the
Type IIA/IIB theory on a ${T^7}/{{Z_2}^n}$ orbifold without (or with)
discrete torsion transforms under a unique $T$-duality transformation
to the IIB/IIA theory on the same orbifold with (or without) discrete
torsion. However, the result of \cite{shat} states that the IIA or
IIB theory on such an orbifold without discrete torsion is equivalent
to the same theory on the orbifold with discrete torsion, up to deformations
in the moduli space. Thus, given one of the Type II theories on
the orbifold without discrete torsion, one can marginally perturb the
theory and one finds the same theory on the same orbifold, but now
with discrete torsion turned on. 

The results we find here are therefore wholly analagous to
the results concerning mirror symmetry for Calabi-Yau ${T^6}/{[Z_2{\times}Z_2]}$
orbifolds \cite{dt}. It is in this sense that PT conjecture
as stated in \cite{pap} is extended for the orbifold cases we consider here to include discrete torsion. One finds that
the dual theory is in fact a marginal perturbation
of the theory one expected from the conjecture in \cite{pap}.

The above comments also beg 
the question: Is there any relationship between the PT
conjecture and the SV conjecture? 
For the class of Joyce compactifications that we consider here, we provide an answer
to this question. It turns out that the SV and PT
conjectures are in fact the same. Thus the $T$-duality between IIA and IIB 
string theories on ${T^7}/{{Z_2}^n}$ orbifolds with and without
discrete torsion give concrete examples of cases for which the SV
generalised mirror conjecture is satisfied.

\section{Dirichlet Joyce Manifolds.}

In \cite{J2} Joyce constructed many examples of compact 7-manifolds
with $G_2$ holonomy as blown up orbifolds of the seven torus, $T^7$. 
The largest class of such manifolds presented in \cite{J2} were
constructed with a ${Z_2}^n$ orbifold group. Of these
$n$ $Z_2$ generators, three act non-freely on $T^7$ (each non-free
$Z_2$ inverting four coordinates). If one compactifies a higher
dimensional locally
supersymmetric theory on such a Joyce space then each non-free
$Z_2$ breaks half of the $N$ supersymmetries which are intact after
compactification on $T^7$, leaving $N/8$. The remaining $n-3$
$Z_2$ generators preserve supersymmetry and act freely on the torus. 
We will
mainly consider ${T^7}/{({Z_2}^3)}$ Joyce orbifolds in this
paper, but in principle, our analysis also applies to the cases with
the additional freely acting generators.

Begin with Type IIA string theory on a $T^7$ with coordinates $x_i$, for
$i=1,2...7$. This theory has $N=16$ supersymmetry in three dimensions.
In accord with our above comments, we can take a ${Z_2}^3$ orbifold 
of this theory which breaks supersymmetry to $N=2$.
The three $Z_2$ generators of the orbifold group, $\Gamma$ 
may be defined as follows \cite{J2}:

\be 
\alpha(x_i) = (-x_1,-x_2,-x_3,-x_4,x_5,x_6,x_7)
\ee
\be
\beta(x_i) = (b_1-x_1,b_2-x_2,x_3,x_4,-x_5,-x_6,x_7)
\ee
\be
\gamma(x_i) = (c_1-x_1,x_2,c_3-x_3,x_4,-x_5,x_6,-x_7)
\ee

The constants, $b_i$ and $c_j$ take values in $(0,1/2)$, and remain to
be specified. Of course the {\it most} general such orbifold can 
have translations in any of the inverted directions for each generator.
This requires specifying four constants for each generator, and
there exist 16 possible choices for each generator. This gives
$16^3$ different orbifolds that one can define. However it is very likely
that there exist large degeneracies between such orbifolds, and 
the number of independent string backgrounds that one can obtain is
probably much smaller. The analysis of the present paper in principle 
applies to all of these cases, but for simplicity we restrict ourselves
to the class of orbifolds defined above. 

Note that 
there are no shifts in the $x_4,x_5,x_6,x_7$ directions. 
Desingularisation of the  ${T^7}/{\Gamma}$ orbifold leads to a compact
7-manifold with holonomy precisely $G_2$, as long as $(b_1,b_2)\neq(0,0)$
and $(c_1,c_3)\neq(0,0)$\ft{For the cases when the constants do not satisfy
both of these constraints, it is at present unknown how to desingularise
the orbifold, although it is still plausible that such a desingularisation
does exist \cite{Joyce}. 
In string theory however there does not seem to be any obstruction
to the definition of the orbifold for any choice of the
constants and one can calculate topological
properties of the compactification space in the standard way. If it does
indeed turn out that such singularities cannot be resolved using classical
geometry, then these cases will be further examples of stringy geometry.}.
In this paper, we restrict ourselves to a study of the cases for which
classical desingularisations are known, although the analysis will also apply
to the other cases.

For notational simplicity, we will denote an element of the orbifold group
which inverts for example the $x_l,x_m,x_n,x_p$ coordinates as 
$I_{lmnp}(a_1,a_2,a_3)$, where the constants $a_1,a_2,a_3$
specify shifts in the $x_1,x_2,x_3$ directions respectively. So for
example, Type IIA theory on the $Z_2$ orbifold defined by $\beta$ will
be denoted by Type IIA on ${T^7}/{I_{1256}(b_1,b_2,0)}$, and so on.

Thus Type IIA string theory on the Joyce orbifold defined above will
be denoted as \nl
Type IIA on \nl
\be
{T^7}/{[I_{1234}(0,0,0), I_{1256}(b_1,b_2,0),I_{1357}(c_1,0,c_3)]}
\ee

The non-zero Betti numbers of a compact 7-manifold of $G_2$ holonomy
are $b_0$=$b_7$=$1$, $b_2$=$b_5$ and $b_3$=$b_4$ (These should not
be confused with the constants $b_i$ above.). The conformal field theory
description of such string backgrounds was given in \cite{shat}(see also 
\cite{jose}). 
$M$-theory
and string theory compactification on such spaces were first studied
in \cite{pap} in the context of duality. 
There it was shown that, after dualising all vectors 
to scalars, the perturbative massless spectrum 
of both the Type IIA and IIB string theories on
such Joyce manifolds consisted of $d=3$, $N=2$ supergravity with
$b_2+b_3+1$ scalar multiplets. Before dualising the vectors, however, the
spectrum in the Type IIA theory is $b_2+1$ vector multiplets and $b_3$ scalar
multiplets. In the ${Z_2}^n$ orbifolds, $b_2$ vector multiplets
and $b_3$ - $7$ scalar multiplets arise in the twisted sectors, with
one vector and $7$ scalar multiplets in the untwisted sector.
Here we will be interested in the $U$-dual
description of such compactifications. This description
will take the form of $D$-manifolds and $M$-manifolds.

Consider first setting the constants $(b_1,b_2,c_1,c_3)$=$(0,1/2,1/2,1/2)$.
In \cite{J2} it was shown that desingularisation of the orbifold defined by
equations $(1)-(3)$ with this choice of constants
gives a Joyce manifold with $b_2=12$ and $b_3=43$. The
Type IIA orbifold description is denoted by:
Type IIA on
\be
{T^7}/{[I_{1234}(000), I_{1256}(0,1/2,0), I_{1357}(1/2,1/2,0)]}
\ee
Now make an $R\rightarrow1/R$ T-duality transformation on the three circles
in the $4,6,7$
directions. This transformation will be denoted by $T_{467}$.
This takes us to an orbifold of Type IIB theory denoted by:
Type IIB on
\be
{{T^7}^{\prime}}/{[I_{1234}(0,0,0).(-)^{F_l},I_{1256}(0,1/2,0).(-)^{F_l},
I_{1357}(1/2,1/2,0).(-)^{F_l}]}
\ee
Here, the prime indicates that this $T^7$ is $T$-dual to the
original one. However, from now on (until section $4$), we will
drop the primes for notational simplicity, with the relationships
between dual circles understood.
The important point to note is that the only non-freely acting members of
the orbifold group are the generators, and that in the IIB description these
are all three of the form $I_{lmno}.(-)^{F_l}$, for some 
$l \neq m$$\neq n$$\neq o$.
It was pointed out in \cite{Kut}, that the ``twisted'' sector states
of such an element in the Type IIB theory consist of sixteen $NS-NS$ 
fivebranes\ft{By ``twisted'' sector we will take to mean
any states which are required for physical consistency of the theory.}. 
In our case, this means we have three sets of such fivebranes, 
which in addition wrap around three sets of three-tori, giving membranes
in three dimensions. Each of these membranes carries one vector multiplet
and three scalar multiplets of the $N=2$ supersymmetry.
However, for each set of fivebranes, we must consider the
action of the orbifold group, which in this case identifies each set in four
groups of four, leaving four independent fivebranes associated with each of 
the three generators. This gives a total of $12$ vector multiplets and
$36$ scalar multiplets coming from the twisted sectors of the orbifold.
From the untwisted sector, we find $1$ vector and $7$
additional scalar multiplets giving
a total of $13$ vector and $43$ scalar multiplets. This is precisely what
we found in the original perturbative Type IIA orbifold. Note that there is
a perfect correspondence between untwisted and twisted sector states in both
descriptions.

We can now use the Type IIB S-duality \cite{U} $Z_2$ element, which
exchanges all objects carrying $NS-NS$ charges with those carrying $RR$ 
charges\ft{It also transforms $(-)^{F_l}$ into $\w$.}. This means we have a 
$D$-strings on $D$-manifold description \cite{D}
of the original Type IIA compactification. The two descriptions are related
by $U$-duality. In particular, the blowing up modes which desingularise the
Joyce orbifold are mapped into wrapped $RR$
fivebrane states in the $D$-manifold
description. The ``addition'' to the (co)homology of the Joyce
manifold which arises from desingularisation in the IIA
description is beautifully encapsulated
by the wrapped fivebrane worldvolume fields in the $D$-manifold description. 

Let us consider a slightly more complicated example. 
This will involve discrete torsion. Set $(b_1,b_2,c_1,c_3)$
=$(0,1/2,1/2,0)$. Begin with Type IIA on the Joyce orbifold defined
by equation $(4)$, with this choice of shift vectors. In the untwisted sector
we find $1$ vector multiplet and $7$ scalar multiplets. The two sets
of sixteen 3-tori
fixed by $I_{1234}(0,0,0)$ and $I_{1256}(0,1/2,0)$ respectively are both
identified in four groups of four by the orbifold group, giving a total
of eight independent contributions to the singular set of the orbifold, which
are locally of the form:
\be
{{R^4}/{Z_2}}{\times}{T^3}
\ee
where the $Z_2$ reflects all four coordinates of $R^4$. Desingularising each 
of these contributes $1$ to $b_2$ and $3$ to $b_3$ of the manifold \cite{J2}.
Desingularisation of the third non-freely acting element, 
$I_{1357}(1/2,0,0)$, is more complicated
and involves discrete torsion \cite{shat}. This is due to the following.
The element ${I_{1234}(0,0,0)}.{I_{1256}(0,1/2,0)}={I_{3456}(0,1/2,0)}$ acts
trivially on the set of sixteen fixed 3-tori of $I_{1357}(1/2,0,0)$. The
element $I_{1256}(0,1/2,0)$ exchanges these sixteen elements in eight pairs
leaving eight singularities associated with $I_{1357}(1/2,0,0)$, each of
which is locally of the form:
\be
{({R^4}/{Z_2}{\times}{T^3})/{{Z_2}^\prime}}
\ee
Here, ${Z_2}^{\prime}$ denotes the action of $I_{3456}(0,1/2,0)$. It turns
out, that such a singularity admits two topologically distinct resolutions,
the details of which may be found in \cite{J2}. This is discrete torsion.
One choice of resolution contributes $1$ to $b_2$ and $1$ to $b_3$. The
other choice contributes $2$ to $b_3$ only. From the Type IIA
orbifold point of view, these two choices of resolution correspond
to a choice in an overall phase factor acting on the twisted sector states
\cite{shat}. A detailed study of orbifolds with discrete torsion
was made in \cite{dt}. 

Of the eight singularities which admit two resolutions, 
if we choose $l$ to admit the first resolution and $8-l$
to admit the second, then all in all the non-trivial Betti numbers of the
Joyce manifold are $b_2=8+l$ and $b_3=47-l$ \cite{J2}. This means, that in
the Type IIA orbifold compactification we will find $1$ vector multiplet
and $7$ scalar multiplets from the untwisted sector plus $8+l$ vector
multiplets and $40-l$ scalar multiplets from the twisted sectors.

As we did in the last example, let us map this Type IIA orbifold
to a Type IIB $D$-manifold. This can be achieved using exactly the same
duality transformations as we considered before. After applying
these we find a Type IIB compactification denoted by:
Type IIB on
\be
{T^7}/{[I_{1234}(0,0,0).\w,I_{1256}(0,1/2,0).\w,I_{1357}(1/2,0,0).\w]}
\ee
Again, the ``twisted'' sectors associated with each of the non freely
acting elements are $16$ $RR$ fivebranes. The first two generators both
contribute, as before, four membranes each. 

However, when the fivebranes associated with the third generator are at
its respective fixed points, the element
${I_{1234}(0,0,0).\w}.{I_{1256}(0,1/2,0).\w}=I_{3456}(0,1/2,0)$ preserves
these fixed points. Therefore, as in \cite{Go}, the action of this
group element
on these fivebranes (there are eight 
independent fivebranes associated with the third generator) must be taken
into account. These
eight fivebranes are the $U$-duals of the eight twisted
sector multiplets in the discrete torsion sector of the perturbative
Type IIA orbifold. 
The four scalar fields on the fivebranes which
interest us here represent the location of the fivebrane in the 
$(x_1,x_3,x_5,x_7)$ directions. 
The action of $I_{3456}(0,1/2,0)$ inverts two of these coordinates, 
$(x_3,x_5)$. We therefore expect that two of the scalars to be
odd under this action. These fivebranes are wrapped around three-tori with
coordinates $(x_2,x_4,x_6)$, and $I_{3456}(0,1/2,0)$ inverts two of
these, $(x_4,x_6)$ as well as shifting $x_2$ by $1/2$. We therefore expect
that two degrees of freedom of the fivebrane vector field will be projected
out. This choice of $Z_2$ action means that each of these eight fivebranes
will contribute $1$ vector multiplet and $1$ scalar multiplet to
the $3d$ $N=2$ theory. These multiplets correspond to the first
choice of resolution (which added $1$ to $b_2$ and $b_3$) of the
singularity in equation $(8)$ in the Type IIA dual.

However, in the Type IIA case we noted that there existed an overall phase
choice of $^+_-1$ corresponding to discrete torsion. Under the $U$-duality
element which maps the Type IIA description to the $D$-manifold configuration,
a consistent picture implies that a second choice of $Z_2$ action on
these fivebrane fields is also possible here. Consequently, this means
inserting an overall minus sign to the action of the $Z_2$, which means
that the fivebrane fields which were odd under our first choice are now
even, and vice-versa. 
With this latter choice of $Z_2$ action, each
of these eight fivebranes will contribute $2$ scalar multiplets to
the $3d$ theory. We will assume that the ambiguity in the choice of
$Z_2$ actions is consistent\ft{Presumably, this can be shown using
standard $D$-brane techniques.}. With this assumption we can 
choose $l$ of these fivebranes to be of the first type and $8-l$ to be
of the latter type.
If we now collect together all contributions to the massless field content
we find a total of $9+l$ vector multiplets and $47-l$ scalar multiplets in
the three dimensional theory. This is precisely what we found in the
perturbative Type IIA desciption of the theory. 

Finally, it is straightforward to see that when configurations of the
above $D$-branes coincide, one can find points in the moduli space
with $U(n)$ gauge symmetry \cite{enhance}.
From the original perturbative Type IIA
orbifold point of view this should presumably be interpreted as
an appropriate $D$-brane wrapping around a vanishing supersymmetric cycle.
In fact, since manifolds of $G_2$ holonomy have only 3-cycles and 4-cycles
which are supersymmetric (see for instance the sixth reference in [9]).
this implies that the $4$-brane in Type IIA
is wrapping around a vanishing 4-cycle.

\section{$M$-manifold Description.}
In this section, we will study Type IIB theory on the Joyce $7$-orbifolds
defined in equations (1)-(3). We will show that this is related to a
certain $M$-manifold configuration of fivebranes on an orientifold of
$M$-theory. This construction is the generalisation to Joyce manifolds
of the $M$-manifold description of Type IIB on $K3$ \cite{w2}. (The
relationship between Type IIB on $K3$ and $M$-theory
was also discussed in \cite{mu}).

We begin then with Type IIB on a Joyce $7$-orbifold. This is defined as:
Type IIB on:
\be
{T^7}/{[I_{1234}(0,0,0),I_{1256}(b_1,b_2,0),I_{1357}(c_1,0,c_3)]}
\ee

As we mentioned earlier, in \cite{pap} it was shown that IIB theory
compactified on a Joyce $7$-manifold gives, after dualising all vectors
to scalars, $b_2$+$b_3$+$1$ scalar $N=2$ multiplets in three dimensions.
However, before dualising the vectors, we find $b_2$+$1$ scalar multiplets
and $b_3$ vector multiplets. The vector fields in the vector multiplets,
come from expanding the ten-dimensional four-form potential of the IIB theory
in a basis of harmonic three-forms on the manifold. (The dual of this
statement is the expansion of the four-form in terms of harmonic four-forms
on the manifold, giving $b_4$=$b_3$ scalar multiplets.)
If we contrast this with the Type IIA
case, we see that the number of vector multiplets in one theory is
the same as the number of scalar multiplets of the other theory, and 
vice-versa. This
is strikingly similar to mirror symmetry for Calabi-Yau threefolds.
On this basis
it would at first sight appear natural to conjecture a ``mirror'' symmetry
for manifolds of $G_2$ holonomy, under which $b_{2} \leftrightarrow b_3$, 
and Type IIA is interchanged with Type IIB. However, in three dimensions
there is the added twist that vectors are dual to scalars, and that
by dualising all vectors, the number of scalar multiplets in both
the IIA and IIB theories is the same, and therefore such 
a conjectured symmetry may not be so straightforward. Because of this it is
still natural to interchange vector multiplets with scalar multiplets, but
with the possibility that $b_2$ and $b_3$ remain invariant ie the manifold
is the same on both sides of the duality map. 
Our results in the next section
will strongly favour against a symmetry which interchanges $b_2$ with
$b_3$, and as we mentioned, will prove the PT and SV
conjectures, which in these cases essentially leave the Betti numbers inert\ft{
The reader is referred to the last section of this paper for the precise meaning of this statement concerning the Betti numbers.}.
A more detailed analysis, in particular quantum, non-perturbative 
aspects of the duality between IIA and IIB theory on the
same manifold of $G_2$ holonomy will appear elsewhere \cite{prog}.

Again, the important point to note about the above
orbifold of Type IIB is that the only non-freely acting
members of the orbifold group are the generators, and that these all
three invert four coordinates of the torus. It then follows from the
work of \cite{w2,sen1}, that the above orbifold of Type IIB theory
on $T^7$ has a description in $M$-theory on $T^8$. We can see this
more explicitly as follows. Make the $T$-duality transformation
$T_{467}$ on the Type IIB orbifold above. 
This maps
us to 
Type IIA on
\be
{T^7}/{[I_{1234}(0,0,0).(-)^{F_l},I_{1256}(b_1,b_2,0).(-)^{F_l},
I_{1357}(c_1,0,c_3).(-)^{F_l}]}
\ee

Now consider $M$-theory on ${T^8}={T^7}\times{S^1}$, 
where this $T^7$ is the same
as that in the orbifold in equation $(11)$ 
and the coordinate labelling $S^1$ is $x_8$.
Then, because of the relationship between Type IIA and $M$-theory \cite{w1}
and a result given in \cite{sen1}, the above orbifold ($(11)$) 
of the
Type IIA theory (and therefore by $T$-duality, the Type IIB 
Joyce orbifold ($(10)$),
is equivalent to the following orientifold of $M$-theory:
$M$-theory on
\be
{{T^7}\times{S^1}}/{[I_{12348}(0,0,0),I_{12568}(b_1,b_2,0),
I_{13578}(c_1,0,c_3)]}
\ee

The non-freely acting elements are the generators, and all three 
are of the form considered in \cite{w2}.
The twisted sector associated with each generator consists therefore of
sixteen $M$-fivebranes\ft{It is natural to impose that the
configuration of fivebranes associated with each $Z_2$ is that
proposed in \cite{w2} in order to achieve anomaly cancellation}, 
which wrap around three-tori. Again, one must
project onto states invariant under the orbifold group. 
For the examples of Joyce manifolds we considered in the Type IIA case,
it is straightforward to check that the massless spectra
in the corresponding $M$-manifold description that we give here are
precisely what we expected from its dual description as a perturbative
Type IIB compactification on a Joyce $7$-manifold, namely $b_2$+$1$
scalar multiplets and $b_3$ vector multiplets
\ft{In the example with discrete
torsion, one must again make the additional assumption (as in 
\cite{Go}), that there are two consistent choices for the 
appropriate $Z_2$ action.}. We do not repeat this
here for brevity, but this can be calculated along
similar lines as we presented in the $D$-manifold description
of Type IIA on a Joyce orbifold.
This gives the $M$-manifold description of
Type IIB theory on these Joyce orbifolds. Shrinking the circle in the
$M$-theory compactification, gives the weak coupling limit of the Type IIB
compactification, as in the $K3$ case \cite{w2}. 

The fact that the massless
spectrum is the same, after appropriate duality transformations, 
as that found in the Type IIA compactification
does not come as a surprise, because it was pointed out in \cite{pap}
that both the Type IIA and IIB theories compactified on the same
Joyce $7$-manifold give the same massless spectrum. 

Finally, as we discussed at the end of the last section, we can
make some general comments about special points in the moduli space
of the above compactifications. Firstly, it was shown in \cite{w2}, that
when two $M$-fivebranes coalesce one gets a non-critical string theory,
which compactified on $S^1$ gives a massless $SU(2)$ vector multiplet of
(the equivalent of) $N=4$ supersymmetry in four dimensions. With more
complicated configurations of fivebranes, this statement has generalisations
to include $A-D-E$ groups \cite{w2}. Moreover, the string theory one gets
is equivalent to that obtained by wrapping a $D$-threebrane of IIB theory
on a vanishing $S^2$ \cite{w3,w2}. It is clear from these comments that in
the $M$-manifolds above, that at special points in the moduli space when
fivebranes coincide, one will find precisely the string theory discussed
in \cite{w2}. However, this theory will be compactified on $T^3$, giving
an enhancement of gauge symmetry in the full three dimensional theory.

From the original perturbative IIB orbifold point of view, this enhancement
of gauge symmetry can come {\it only} from a threebrane wrapped around
an ${S^2}{\times}{S^1}$ submanifold of ${S^2}{\times}{T^3}$, where the $S^2$ is
shrinking to zero size. The number and singularity structure of such
vanishing 3-cycles which arise from desingularising the orbifold 
is in one-to-one correspondence with similarly
obtained vanishing 4-cycles in the Type IIA case of the previous section.
This is in accord with the PT conjecture.

\section{The PT and SV Conjectures.}

In \cite{pap}, it was conjectured that
the Type IIA and Type IIB theories on the {\it same} Joyce $7$-manifold
are physically equivalent. We will show in this section that this is indeed
true up to moduli deformations.
In fact,
the two are related by $T$-duality, as one might have anticipated.

In section two, we described the $D$-manifold description of Type IIA
theory on a class of Joyce $7$-manifolds. This was essentially a 
generalisation of the $D$-manifold description of Type IIA on
$K3$ \cite{Kut,Op,D}. In the last section, we described the $M$-manifold
description of Type IIB on the same class of Joyce manifolds. This
description was essentially a generalisation of Wittens $M$-manifold
construction \cite{w2}, 
to the case of $G_2$ holonomy. The fact that all these descriptions give
the same low energy spectrum is further evidence that the conjecture of
\cite{pap} is correct. 

In the Type IIA compactification on the Joyce orbifolds, $J$,
that we have considered here, we gave three different $U$-dual descriptions:
$(i):$ Type IIA on $J$.
$(ii):$ $M$-theory on $J\times{S^1}$.
$(iii):$ Type IIB $D$-manifold.

For Type IIB on $J$ we have:
$(i):$ Type IIB on $J$.
$(ii):$ $M$-theory $M$-manifold.

In order to show that all five of these descriptions are equivalent
under duality transformations between expectation values
of some of the moduli fields, it is sufficient to show that any of
the first three are dual to any of the second two; the further equivalences should
follow from the duality transformations we have used in previous sections.
We will show, that for the class of Joyce manifolds that
we have studied in this paper, Type IIA on $J$ is $T$-dual to Type IIB on
$J^{\prime}$, where the difference between $J$ and $J^{\prime}$ is
given by $R\rightarrow1/R$ transformations on some of the circles
in the original $T^7$; a further difference will be that
if the theory on $J$ has no discrete torsion, then $J^{\prime}$ will
have discrete torsion, and vice-versa. 

Consider our original Type IIA compactification, equation $(4)$.
If we make the $T$-duality transformation $T_{235}$ on this compactification 
we end up with the following:
Type IIB on
\be
{{T^7}^{\prime}}/{[I_{1234}(0,0,0),I_{1256}(b_1,b_2,0),I_{1357}(c_1,0,c_3)]}
\ee
Here, we have reinstated the prime to indicate the $T$-duality map between the
compactification moduli in the two theories.
At first sight this, 
by definition, is the Type IIB theory on the ($T$-dual of the)
Joyce manifold we
started with. We would have therefore shown, that for this class of Joyce
manifolds, $T$-duality interchanges the Type IIA and Type IIB theories
compactified on the same manifold. For this class of Joyce $7$-manifolds,
this would give a simple proof of the 
PT conjecture \cite{pap}. However, there is a subtlety
involved here. Orbifold conformal field theories are not
unique \cite{dt1}. In general it is possible to introduce discrete
phases in the path integral twisted sectors under the
constraints of modular invariance \cite{dt1}.
In the above $T$-duality transformation between the two 
theories, additional phases in the path integral of these theories will
not be transparent. These phases, if present, correspond to discrete torsion
\cite{dt1}. 

The relationship between $T$-duality, mirror symmetry and
orbifolds with discrete torsion was investigated in detail in \cite{dt}.
There it was found that for the Type IIA/B theory compactified on a Calabi-Yau
${T^6}/[Z_2{\times}Z_2]$ orbifold without discrete torsion, 
the mirror theory is a $T$-dual compactification of the Type IIB/A theory
on the orbifold with discrete torsion, and vice versa. This equivalence
was proved to genus $g$ in \cite{dt}. Given that our orbifolds here contain
just an extra $Z_2$ generator, one would expect some analagous $T$-dual
relationship here. 
The only other difference technically, 
between the orbifolds here
and that considered in \cite{dt} is that our orbifold generators induce
extra translations of the coordinates of the torus.
Consider, for example, the following Calabi-Yau compactification:
Type IIA/IIB on
\be
{T^6}/{[I_{1234}(0,0,0),I_{1256}(1/2,0,0)]}
\ee
The only difference between this case and that considered in \cite{dt}
is the presence of a non-zero translation of the coordinate in the $x_1$
direction. The only non-freely acting elements of the orbifold group
are the two generators. The singular set of the first generator is
sixteen copies of $T^2$. However, these are interchanged in eight pairs
by the second generator, leaving eight invariant $T^2$'s coming from
this element. The same is true for the second generator. Its sixteen
fixed tori are reduced to eight by the action of the other $Z_2$.
Thus altogether, we have sixteen $T^2$'s even under the action of $Z_2{\times}
Z_2$ and sixteen which are odd. $(h_{11},h_{21})$ in the untwisted sector
are $(3,3)$. Each even $T^2$ adds one to both of these, giving $(19,19)$
for the final Hodge numbers.

Now, let us turn on discrete torsion. This introduces an overall phase
factor of $-1$ in the twisted sectors, and consequently means that states
which were even(odd) are now odd(even). In this case, because the states
which were odd without discrete torsion also constitute sixteen tori, the
end result is to produce a conformal field theory which also computes 
$(19,19)$ for the Hodge numbers. This is of course expected as $h_{11}$
= $h_{21}$ for the original orbifold. The result of \cite{dt} also applies
to this case, and the case without discrete torsion is $T$-dual to
the case with discrete torsion, and vice versa. The requisite
transformation in this case is $T_{246}$. Further, the $T$-duality
transformation is the mirror transformation in this case also.

Thus, formally
we may conclude that the Type IIA/IIB compactification $(14)$ without
discrete torsion is mirror to the IIB/IIA compactification $(14)$ with
discrete torsion. We now move on to discuss the compactifications which
have been the focus of this paper.

Above we showed that, under the transformation $T_{235}$, the orbifold
generators of our original Type IIA compactification (without discrete
torsion), transform into the same generators of a Type IIB compactification.
We have not yet checked whether or not this IIB compactification has
discrete torsion turned on. All that is required now is a straightforward
generalisation of the computation in \cite{dt}, of how the path integral
for the theory transforms under $T_{235}$. As pointed out in \cite{dt}, this
can be done by computing the transformation of the path integral measure.
For definiteness we will restrict ourselves to our
example orbifold which had $(b_1,b_2,c_1,c_3)=(0,1/2,1/2,0)$,
but the result is independent of the shift
vectors in equations (2) - (3).

Consider again then this Type IIA orbifold {\it without} discrete torsion.
(The Betti numbers for the classical manifold in this case are $b_2$=$8+l$
and $b_3$=$47-l$, with $l=0,1,..8$). The orbifold isometry group
is ${(Z_2)^3}$, so the discrete
torsion group is ${(Z_2)^2}$ \cite{dt1}, as for each element, there is a
$Z_{2}{\times}Z_2$ choice for the overall phase in the path integral 
associated with the other generators of the group. As we discussed earlier
using fivebrane twisted sectors, in this example (without discrete torsion),
the singular set associated with each of three non-freely acting elements
is sixteen three-tori, each of which get interchanged by the remaining
$Z_2{\times}Z_2$ group. The first and second generators contribute
four independent three-tori each.
However as we discussed,
the third generator contributes a further
eight three-tori which are however acted on by a further $Z_2$ (see
equation (8)). All in all
this means that the Betti numbers of the orbifold with no discrete torsion
are $b_2=16$ and $b_3=39$.

Turning on discrete torsion in the
string path integral simply means inserting an overall minus sign
in the twisted sector associated with each of these three elements when
projecting onto group invariant states. (We give the explicit form of the
discrete torsion in equation (17) below.)

Here (using the notation of
equations $(1)-(3)$), if we consider the twisted
sector associated with $\alpha$, we have to project onto states invariant
under the group generated by $\beta$ and $\gamma$. In this case, we are
free to insert an overall minus sign in the action of $\beta$, $\gamma$ and
$\beta\gamma$ on the $\alpha$ twisted sector. This phase is discrete torsion.
There exist analogous choices in the remaining two twisted sectors.
In the example we are considering here, it is straightforward to check
that in the twisted sectors corresponding to the first two generators,
all three choices for the torsion give the same contribution to the
Betti numbers
as the case without discrete torsion. For the eight three-tori associated with
the third generator in this example, turning on discrete torsion means that
states which were odd under the $Z_2$ in equation $(8)$ in the case without
the torsion are now even and vice versa. Geometrically this corresponds to the
fact that these singularities admit two topologically distinct resolutions
\cite{J2,shat}. For us it means that the orbifold with discrete torsion has
Betti numbers $b_2=8$ and $b_3=47$.

Therefore if we could prove that the transformation $T_{235}$ as
well as converting IIA to IIB also turns on discrete torsion, then we
would have proved in this case our extended PT conjecture.

If we begin with Type IIA theory on
the orbifold without any discrete torsion then we have Type IIA
on the Joyce manifold with $b_2=16$ and $b_3=39$. After making the
transformation $T_{235}$ we would then end up with Type IIB on the 
Joyce manifold with $b_2=8$ and $b_3=47$. However, as shown in \cite{shat},
in string theory one can smoothly interpolate between any of the 
manifolds in this family (ie any value for $l$) by turning on 
marginal perturbations. Thus at the level of conformal field
theory the backgrounds on the IIA and IIB sides are equivalent.
Thus if we could prove that the IIA theory without discrete torsion
is equivalent to the IIB theory with discrete torsion, then we
have essentially proven the PT conjecture.

We will now
complete the proof of the PT conjecture along the lines of the
computation in \cite{dt}.

\subsection{Genus $g$ Transformation of Path Integral.}

A genus $g$ Riemann surface, ${\Sigma}^g$, has $2^{2g}$ spin structures.
If the number of fermion zero modes of a given chirality is even or odd,
then a given spin structure, $\alpha$ is said to be even or odd.
It is a well known fact that under an $R\rightarrow1/R$ transformations
on circles in, say, the $x_i$ directions, the left-moving fermions which
partner the string coordinates in those directions pick up
a minus sign. In fact this is one of the fundamental reasons why the Type IIA
and IIB theories are related by $T$-duality. The non-zero modes of these
fermions are naturally paired, so these modes will not contribute any
change in sign to the measure of the path integral. Thus, under such
$T$-duality transformations, the transformation of the measure of the path
integral is determined by the number of fermion zero-modes in the
transformed directions being even or odd. Thus, for spin structure $\alpha$,
the measure, $\mu$, of the genus $g$ path integral of a toroidal target space
transforms as \cite{dt}

\be
{\mu}_{g,\alpha} \rightarrow {(-)^{{\sigma}_{\alpha}}{\mu}_{g,\alpha}}
\ee

where ${\sigma}_{\alpha}$ is $0$ or $1$ if the spin structure $\alpha$ is
even or odd. The above formula asserts that the IIA/IIB theory on a torus 
transforms into the IIB/IIA theory on the $T$-dual torus, under an odd number
of $R \rightarrow 1/R$ transformations. Before discussing the orbifold
case, we will briefly define the spin structure parity,
$\sigma_{\al}$ for a genus $g$ Riemann surface, $\Sigma^g$. This was
discussed in \cite{dt}.

On ${\Sigma}^g$, we can choose a canonical basis of $1$-cycles, $(a_i,b_i)$,
for $i=1,..g$. This basis defines a canonical spin structure. Thus, we can
specify a spin structure $\alpha = (\theta_i, \phi_j) \equiv (\Theta, \Phi)$
by the $Z_2$ valued quantities $\Theta$ and $\Phi$ ($(\Theta,\Phi)$ are 
$g$-vectors with components $(\theta_i,\phi_i)$). In the spin structure
$\alpha$, fermions have an extra minus sign in their $(a_i,b_i)$
boundary conditions,
relative to the canonical spin structure, such that the components of $\Theta$
or $\Phi$ are one \cite{dt}. The parity of the genus $g$ spin structure
is
\be
{\sigma}_{\alpha}\equiv{\sg}(\Th,\Ph)=\Theta.\Phi \ \  mod\  2
\ee

Now we discuss the modification to the above transformation for the
toroidal orbifolds we have been discussing in this paper. Our
orbifolds are of the form ${T^7}/{\Gamma}$, with $\Gamma$ the
${Z_2}^3$ group generated by $\alpha,\beta,\gamma$ of equations $(1)-(3)$.
We must therefore consider the $\Gamma$ twists around the $a_i$ and $b_i$
directions. 
In the $(a_i,b_i)$ directions, a general $\Gamma$ twist will be of the
form $({\al}^{r_i}{\bt}^{s_i}{\ga}^{t_i},{\al}^{u_i}{\bt}^{v_i}{\ga}^{w_i})$, 
or
$({\al}^{R}{\bt}^{S}{\ga}^{T},{\al}^{U}{\bt}^{V}{\ga}^{W})$ 
in $g$-vector notation.

The discrete torsion for this theory is given by
\be
\epsilon = (-1)^{R.V-S.U+R.W-T.U+S.W-T.V}
\ee

This formula is the generalisation to the ${Z_2}^3$ case of the ${Z_2}^2$
case given in \cite{dt}. A simple way to see this is that the above discrete
torsion should reduce to that considered in \cite{dt}, when any one of the
three $Z_2$ generators acts trivially; ie consider the twists when any single pair of g-vectors, $(R,U), (S,V), (T,W)$ is zero. When this is the case, the twist reduces to a $Z_2 \times Z_2$ and therefore the formula should reproduce that of \cite{dt}. Thus the above formula is the
unique generalisation of the one given in \cite{dt}. In fact the formula above
is just three copies of that in \cite{dt}, one for
each $Z_2{\times}Z_2$ subgroup of $\Gamma$.

Having calculated the discrete torsion phase factor for the Joyce orbifolds
which have been our interest, our aim is to show that, under $T_{235}$,the
path integral measure at genus $g$, with spin structure $\al$ transforms as

\be
{\mu}_{g,\al} \rightarrow {(-)^{{\sigma}_{\al}}}.\epsilon.{\mu}_{g,\al}
\ee
with $\sigma$ and $\epsilon$ as defined in equations $(16)$ and $(17)$.
If this formula is true, then the IIA/IIB theory without (or with) discrete
torsion transforms under $T_{235}$ into the IIB/IIA theory with (or without)
discrete torsion. The entire transformation would be a consequence of the
fact that left-moving zero modes in the $T$-dualised directions change
sign. We can view the transformation as follows. The first factor
converts IIA/IIB into IIB/IIA, while the second turns discrete torsion
on (or off), if the undualised theory had discrete torsion turned off
(or on) \cite{dt}.

Since we are inverting radii in the $(2,3,5)$ directions, and it is the
fermions (which superpartner the string coordinates in these 
directions) which change sign in the toroidal compactification, we need to
identify the ``twisting'' of a toroidal spin structure in these directions, 
say $\al$=$(\Theta,\Phi)$ when twisted by the orbifold group. By inspecting
the action of $\al,\bt,\ga$ on these coordinates, and with the discrete
torsion defined above, it is easy to see that the spin structure, 
$(\Theta,\Phi)$ is shifted in these directions to the following:

\be
x_2:\ \ \ (\Theta+R+S, \Phi+U+V)
\ee
\be
x_3:\ \ \ (\Theta+R+T, \Phi+U+W)
\ee
\be
x_5:\ \ \ (\Theta+S+T, \Phi+V+W)
\ee

In the orbifold theory, the path integral transforms as follows:\nl
\be
{\mu}_{g,\al}\rt{(-)^{[{\sg}(\Th+R+S,\Ph+U+V)+{\sg}(\Th+R+T,\Ph+U+W)+
{\sg}(\Th+S+T,\Ph+V+W)]}}.{\mu}_{g,\al}
\ee \nl

Using $(16)$ and $(17)$, or Theorem 2 of \cite{Atiyah} (which is
also stated in \cite{dt}) which states:
\be
{\sg}(\Th+R+U,\Ph+S+T)\ =\ {\sg}(\Th,\Ph)+{\sg}(\Th+R,\Ph+S)+
{\sg}(\Th+U,\Ph+T)+(R.T-S.U)
\ee
it is straightforward to verify that the above
transformation is precisely the required transformation $(18)$.
This completes the proof that the IIA/IIB theory on the ${T^7}/{\Gamma}$
orbifold without (or with) discrete torsion is equivalent to the IIB/IIA
theory on the ${T^7}^{\prime}/{\Gamma}$ orbifold 
with (or without) discrete torsion, where the prime denotes that the two
$7$-tori are related by $T_{235}$. We now discuss the implications of this
proof for the SV conjecture.

\subsection{Relation to SV Conjecture.}

Up to this point we have had very little to say about the SV generalised
mirror conjecture \cite{shat}. We will now show that our proof
of equation $(18)$ and our discussions above in fact tell us that the
SV conjecture and the PT conjecture are
in fact the same conjecture for the cases to which we have restricted 
ourselves. In fact, being much more general, the SV conjecture turns
out to encompass the PT conjecture, with the latter a special case of the
former. By establishing this fact momentarily, our proof of $(18)$ also
yields concrete examples of cases in which the SV conjecture applies. 

The reason that the conjectures coincide for the compactifications
which have interested us here is the following. Under the duality 
transformation, $T_{235}$ which interchanges the IIA and IIB theories, 
the operator, $(-)^{F_l}$, changes sign. This means that $(-)^F$ also
changes sign. This is precisely the kind of situation to which
the SV generalised mirror conjecture applies. As we have 
shown that the two theories are exactly equivalent
under $T_{235}$, and also that this proves the PT conjecture for these cases,
the SV and PT conjectures for IIA/IIB compactification on manifolds
of $G_2$ holonomy are in fact the same.

\section{Discussion.}

We have discussed in this paper perhaps the simplest class of
Type II vacua in three dimensions which have $N=2$ supersymmetry.
We hope we have convinced the reader that there is a rich structure
in this class of vacua which deserves to be further explored.
One interesting avenue for this is the following. The strong coupling
limit of a $d=3$, $N=2$ Type IIA theory, obtained by compactifying
on some ``$G_2$'' background, $J$, is a $d=4$, $N=1$ vacuum obtained
by compactifying $M$-theory on $J$. It would certainly be interesting to
study which kind of physics this will lead to. In fact since manifolds of
$G_2$ holonomy contain only supersymmetric 3-cycles and 4-cycles, $M$-theory
on $J$ can have supersymmetric membrane instantons which come from wrapping
membranes on such 3-cycles. It is plausible that these could generate a non-perturbative superpotential via a mechanism discussed in \cite{w4}.
The magnetic duals of these objects are
strings which come from wrapping fivebranes on 4-cycles. Moreover, when
such 4-cycles vanish, one should get an interesting tensionless
string appearing in four dimensions. Tensionless strings in four dimensions
have been discussed
recently in \cite{kleb}.

For this same simple class of vacua, we have also given a proof of the
PT and SV ``mirror'' conjectures using $T$-duality. This, we
believe puts these ``mirror'' conjectures on a much firmer footing.
We hope that this
convinces the reader that the geometry and physics of more general ``$G_2$''
vacua is possibly as interesting as the Type II Calabi-Yau ``mirror vacua''. 

\section{Acknowledgements.}

We would like to thank Jose Figueroa-O'Farrill, Jerome Gauntlett, Chris Hull,
Dominic Joyce and Bas Peeters for discussions. We would also like to
acknowledge the hospitality of the organisers of the Network Meeting on 
Field Theory and Strings, Heraklion, Crete (9-14th September 1996) and also the
High Energy Physics Group, ICTP, Trieste, where these results were first
presented. Finally we thank PPARC, by whom this work is supported.

\end{document}